\documentclass{wqcd03} 
\usepackage{mathptmx}                 
  
\newcommand{\be}{\begin{equation}}
\newcommand{\ee}{\end{equation}}
\newcommand{\bea}{\begin{eqnarray}}
\newcommand{\eea}{\end{eqnarray}}

\confname{QCD@Work 2003 - International Workshop on QCD, Conversano, Italy, 
14--18 June 2003}

\title{Non factorizable Effetcs in $B \to \chi_{c0} K^- $  from 
Charmed Meson Rescattering\thanks{Speaker at the Workshop.}.}

\author{ T. N. Pham\addressmark{a}}

\address[a] {Centre de Physique Theorique, \\
Centre National de la Recherche Scientifique, UMR 7644, \\  
Ecole Polytechnique, 91128 Palaiseau Cedex, France\\ }

\begin{document}

\begin{abstract} 
The $B^{-} \to \chi_{c0} K^{-}$ decay which has no visible factorizable
amplitude, could be induced by the Cabibbo-allowed, color-favored
$B^{-} \to D_{s}D^{*0}$, $B^{-} \to D^{*0}_{s}D^{0}$ and $B \to D_{s}^{*}D^{*}$  via the
rescattering of these charmed mesons to $\chi_{co} K^{-}$.
In this talk I would like to discuss a recent
calculation \cite{Colangelo} of  these effects for 
$B^{-} \to \chi_{c0} K^{-}$
and $B^{-} \to J/\psi K^{-}$ decays. We find that charmed 
scattering effects seem capable of producing a large 
$B^{-} \to \chi_{c0} K^{-}$ branching ratio measured by the 
Babar and Belle Collaboration and make a significant
contribution to the $B^{-} \to J/\psi K^{-}$ branching ratios in agreement
with experiments.
  
\end{abstract}

\maketitle

\section{Introduction}
Understanding colored-suppressed $B$ decays is a 
challenging problem in $B$ decays. It is well-known that  
colored-suppressed non leptonic $B$ decays 
have a large branching
ratios compared with naive factorization model. The effective 
Wilson coefficient $a_{2} = c_{2} + c_{1}/N_{c}$ is far smaller than the 
experimental value found from the measured branching ratios, like 
$B^{0} \to D^{0}\pi^{0}$, $B^{0} \to D^{0}\rho^{0}$ and also for
$B^{-} \to J/\psi K^{-}$ decays. Improved QCD factorization 
could produce a larger $a_{2} $, but the predicted branching ratios
for these colored-suppressed decays are still below the measured values,
indicating possible nonfactorizable terms. Another evidence could come   
from the decay $B^{-} \to \chi_{c0} K^{-}$ with a large branching ratio
measured by Belle\cite{Belle1} and Babar\cite{Babar} Collaboration:
\bea
&&{\rm BR}(B^-\to \chi_{c0} K^-) = (6.0^{+2.1}_{-1.8})\times 10^{-4}
(\rm Belle) \nonumber\\
&&{\rm BR}(B^-\to \chi_{c0} K^-) = (2.4\pm 0.7)\times 10^{-4} (\rm Babar)
\label{chi0}
\eea
which is comparable to the $B^{-} \to J/\psi K^{-}$ branching ratio 
of $(10.0\pm 0.5)\times 10^{-4}$\cite{PDG}. This is a big surprise since 
there is no appreciable factorizable contribution to this 
decay \cite{Suzuki}. In fact, in the naive factorization model,  
because of the conservation of the vector current
$\bar{c}\gamma^{\mu}c$, the matrix element 
$<0|\bar{c}\gamma^{\mu}c|\chi_{c0}> =0$ and the  decay amplitude
$B^{-} \to \chi_{0}K^{-}$ vanishes.  
The large branching ratios for this decay could be an evidence for
a non factorizable contribution in non leptonic $B$ meson decay with
charmonium in the final state. One possible non factorizable  effects could
be  induced by the Cabibbo-allowed, color-favored $B \to D_{s}D^{*}$ 
and $B \to D^{*}_{s}D^{*}$  decays via 
the rescatterings of charmed mesons into charmnonium and $K$ or $K^{*}$
meson in the final state (inelastic FSI effects). In fact,  the 
Cabibbo-allowed, color-favored $B$ decays to charmed mesons,  like $B \to
D_{s}D^{*}$ and $B \to D^{*}_{s}D^{*}$  etc. with
branching ratios a few times $10^{-2}$ would be the dominant contribution
to the absorptive part of the $B^{-} \to \chi_{c0} K^-$ and 
$B^{-} \to J/\psi K^{-}$ decay amplitudes. This is a rare situation in  
non leptonic $B$ decays, similar to  the $B \to K\pi$ decays for which 
non factorizable contributions have been estimated in recent
work \cite {Isola} in which a large absorptive part and a large CP 
asymmetry for $B \to K\pi$ and  $B \to \pi\pi$ amplitude are obtained.
The idea that long-distance inelastic FSI effects 
could be present in heavy meson decays due to
charmed meson rescattering process has been considered
in the past, for $B \to \pi\pi$ \cite{Kamal}, for 
OZI-suppressed heavy quarkonium decays \cite{Achasov} and
for  $B_{s}\to \gamma\gamma$ decay \cite{Choudhury},  
As mentioned, the large branching ratios for the Cabibbo-allowed, 
color-favored two-body $B$ decays with charmed meson in the 
final state, e.g $B \to D^{*}_{s}D^{*}$ decay, make the rescattering 
effects for $B \to K\pi$ , $B^{-} \to \chi_{c0} K^-$ and 
$B^{-} \to J/\psi K^{-}$ decay more important than for other 
OZI-suppressed heavy quarkonium decays. In other word, charmless $B$ decays
and charmonium $B$ decay with $K$ or $K^{*}$ in the final state
are  favorable decays  to look for inelastic FSI effects. 
In a recent work \cite{Colangelo}, we computed the non factorizable
$D_{s}D^{*0} $ ,  $D^{*}_{s}D^{0} $ and 
$D^{*}_{s}D^{*0} $ rescattering terms for  the $B^{-} \to J/\psi K^{-}$ 
and  $B^{-} \to \chi_{c0} K^-$ decays and obtain 
large branching ratios in more or less agreement with experiments.  
Before presenting the calculations, I would like to discuss the
QCD factorization  for $B^{-} \to J/\psi K^{-}$ decay.

\section{$B^{-} \to J/\psi K^{-}$ in QCD factorization}

The effective Hamiltonian for nonleptonic $B$ decays:
\begin{eqnarray}
&&\kern -0.7cm  {\mathcal H}_{\rm eff}\kern -0.1cm =\kern -0.1cm  \frac{G_F}{{\sqrt 2}} 
[V_{ub} V_{us}^*(c_1 O_{1u} \kern -0.1cm+\kern -0.1cm c_2 O_{2u} ) 
 +  V_{cb} V_{cs}^* (c_1 O_{1c} +c_2 O_{2c} ) ]\nonumber\\
&&\kern -0.7cm - \sum_{i=3}^{10} ([V_{ub} V_{us}^* c_i^u
     +  V_{cb} V_{cs}^* c_i^c + V_{tb} V_{ts}^* c_i^t) O_i ] + \rm h.c.
\label{eq1}
\end{eqnarray}
where the superscripts $u$, $c$, $t$ denote the internal quark.
The tree-level $(V-A)\times (V-A)$ operators are,
\bea
&&O_{1q} = \bar q \gamma_{\mu} (1 - \gamma_5) b
   \bar s \gamma^{\mu} (1 - \gamma_5) q , \nonumber \\
&&O_{2q} = \bar q \gamma_{\mu} (1 - \gamma_5) q
   \bar s \gamma^{\mu} (1 - \gamma_5) b.
\label{O1}
\eea
with $q= u,c$ . For the penguin operators, we rewrite
$O_3$ $-$ $ O_6$, using the Fierz transformations, as follows:
\bea
&&O_3 = \sum_{q = u,d,s,c,b} \bar s \gamma_{\mu} (1 - \gamma_5) b
\bar q \gamma^{\mu} (1 - \gamma_5) q , \nonumber \\
&&O_4 = \sum_{q = u,d,s,c,b} \bar s \gamma_{\mu} (1 - \gamma_5) q
\bar q \gamma^{\mu} (1 - \gamma_5) b, \nonumber \\
&&O_5 = \sum_{q = u,d,s,c,b} \bar s \gamma_{\mu} (1 - \gamma_5) b
\bar q \gamma^{\mu} (1 +  \gamma_5) q , \nonumber \\
&&O_6 = - 2 \sum_{q = u,d,s,c,b} \bar s (1 + \gamma_5) q
\label{O3}
\eea
\begin{figure}[ht]
\hbox to\hsize{\hss
\includegraphics[width=\hsize]{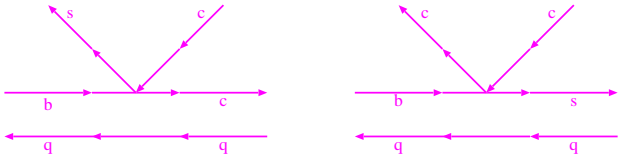}
\hss}
\caption{factorization contribution to \break $B^{-}\to J/\psi K^{-} $ decay }
\label{fig:fact1}
\end{figure}

\begin{figure}[ht]
\hbox to\hsize{\hss
\includegraphics[width=\hsize]{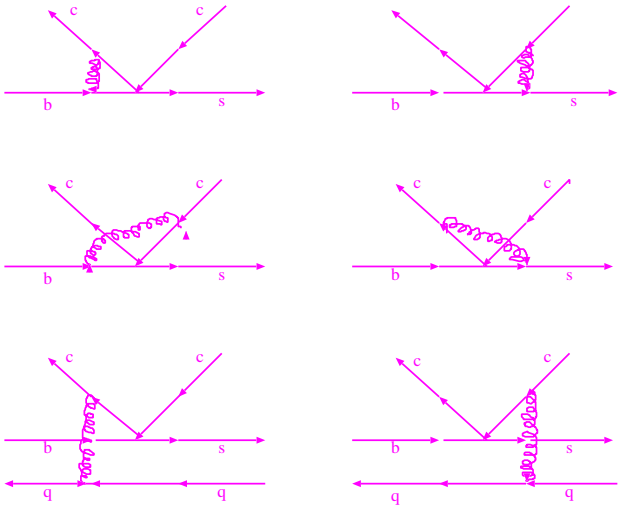}
\hss}
\caption{Vertex and spectator corrections to  $B^{-}\to J/\psi K^{-} $ decay }
\label{fig:fII}
\end{figure}
The factorization approximation (Fig.1) is obtained by neglecting in the
Lagrangian terms which are the product of two
color-octet operators after Fierz reordering of the quark fields. The
effective Lagrangian for non leptonic decays are then given with
$c_{i}$ replaced by $a_{i}$. 
For $N_{c}=3$, $m_b = 5 \rm \, GeV$, we have \cite{Deshpande,Isola1}
\begin{equation}
\begin{array}{rlrl}
a_1=& \ 1.05,  &a_2= & \ 0.07, \\
a_4= & -0.043- 0.016 i, \\
a_6= & -0.054- 0.016 i.\\
\end{array}
\label{ai1}
\end{equation}

Only $O_{2c} = \bar c \gamma_{\mu} (1 - \gamma_5) c
   \bar s \gamma^{\mu} (1 - \gamma_5) b$ contributes to 
$B^{-} \to J/\psi K^{-}$ decay (neglecting penguins):
\bea
 A(B^{-}\to J/\psi K^{-}) 
&=&\,{G_F\over\sqrt{2}}V_{cb}V_{cs}^*(a_2)f_{J/\psi}
m_{J/\psi} \nonumber \\
&&\times F_1^{BK}(m^2_{J/\psi})(2\epsilon^{*}\cdot p_B),  \label{fact}
\eea

With $f_{J/\psi} = 405\pm 15 \rm\,MeV$ , 
$|V_{cb}| = 0.040$ ,$|V_{cs}| = 0.9735$, $F_1^{BK}(m^2_{J/\psi} = 0.70 $ 
one gets a branching ratio 1/10 of the measured value of
$(1.00\pm 0.1)\times 10^{-3}$. 
In terms of $a_{2}$ and $F_{1}^{BK}(m_{J/\psi}^{2})$. the branching ratio
is then
\be
\rm BR(B^{-}\to J/\psi K^{-})= 3.04\times 10^{-2}\,(a_{2}\,F_{1}^{BK}(m_{J/\psi}^{2}))^{2}
\label{br}
\ee
One would need $a_{2}$ in the range $0.25-0.40$ 
depending on the value for the  $B \to K$ form factor 
$F_{1}^{BK}(m_{J/\psi}^{2})$ 
to produce the experimental value. For example, from the
$B^{-}\to J/\psi K^{-} $ branching ratio, one gets $a_{2} =0.38$ 
according to Ref.\cite{Colangelo1}  while in \cite{Cheng} 
an effective  $a_{2} =0.25$ could also give the correct branching ratio.
Non factorizable terms due to vertex correction and spectator 
interactions obtained from QCD factorization
represented by diagrams in Fig.2 have been done for $B^{-}\to J/\psi K^{-} $ 
decay \cite{Chay,Cheng1} who give, in terms of the vertex correction
$f_{I}$ and the hard spectator interaction term $f_{II}$,
\bea
&&a_2 = c_2 +\frac{c_1}{N_{c}} +\frac{\alpha_s}{4\pi} \frac{C_F}{N^{c}} c_1
 \Bigl( -18 +12\ln \frac{m_b}{\mu} \nonumber \\
&& \hspace{3.1cm} + f_I + f_{II} + {\frac {F_{0}^{BK}(m_{J/\psi}^{2})} {F_{1}^{BK}(m_{J/\psi}^{2})}}\,g_{I}
\Bigr)
\label{a2}
\eea
with $f_{I}$, $f_{II}$ and $g_{I}$ given in \cite{Chay,Cheng1} . This result  
shows that $a_{2}$ , as in $B \to \pi\pi$ decays \cite{Beneke}, 
gets a large contribution from  the hard spectator 
interaction term ($f_{II}$) which can increase $a_{2}$ to $0.15-0.20$,  
a significant improvement over the naive factorization model. 
One could vary the form factor $F_1^{BK}(m^2_{J/\psi}) $  to get a bigger 
$a_{2}$ . However a large  $F_1^{BK}(m^2_{J/\psi}) $ would 
imply a large $F_1^{B\pi}(m^2_{\pi})$ and hence a too large 
$B \to \pi^{+}\pi^{0}$ branching ratio. Thus taking the theoretical
uncertainties on the $B \to K$ form factors into account, it seems
that we need a large $a_{2}$  to explain the $B^{-}\to J/\psi K^{-} $  
branching ratio. For $B^{-} \to \chi_{c0} K^{-}$ decay, recent 
work \cite{Song} indicates that there are infrared divergence problems 
in the vertex correction and spectator interaction terms in QCD 
factorization. We now turn to the calculation of the charmed 
meson rescattering effects for $B^{-} \to \chi_{c0} K^{-}$ and 
$B^{-}\to J/\psi K^{-} $ decays.
\section{$B^{-} \to \chi_{c0} K^{-}$ from charmed meson rescattering}
The decay $B^{-} \to \chi_{c0} K^{-}$ can occur through the Cabibbo-allowed,
colored-favored $B^{-}\to D_{s}^{-}D^{*0}$ decay followed by the 
rescattering $D_{s}^{-}D^{*0}\to \chi_{c0} K^{-}$ as well as through the
 $D_{s}^{*-}D^{0}$ and  $D_{s}^{*-}D^{*0}$  intermediate states which
rescatter into $\chi_{c0} K^{-} $ final state.  
\begin{figure}[ht]
\hbox to\hsize{\hss
\includegraphics[width=\hsize]{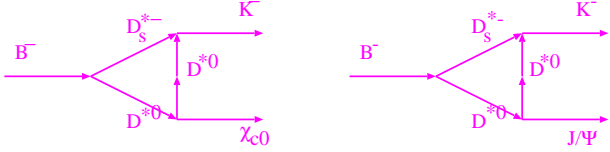}
\hss}
\caption{A rescattering diagram for $B^{-}\to \chi_{c0} K^{-} $ 
and $B^{-}\to J/\psi K^{-} $ }
\label{fig:graph1}
\end{figure}

Experimentally the $B^{-}\to D_{s}^{-}D^{*0}$ decay rate is about a factor
of $20-50$  bigger than the $B^{-} \to \chi_{c0} K^{-}$ so 
 these charmed meson intermediate states give  the dominant contribution
to the absorptive part in the unitarity relation. For the
$D^{*-}_s, D^{*0}$ intermediate state, the absorptive part is given by
\bea
{\rm Im} \, A_1 = {|p_{1}| \over 16 \pi m_B}
\int_{-1}^{+1}\kern -0.5cm &&dz\, A(B^- \to D^{*-}_s  D^{*0})\nonumber \\
\kern -0.5cm &&\times  A(D^{*-}_s  D^{*0} \to \chi_{c0} K^{-})
\label{absorptive}
\eea
where $p_{1}$ is the  of the charmed meson 3-momentum in the rest frame
of the $B$ meson. 
The  amplitude $ A(B^- \to D^{*-}_s  D^{*0}) $~ can be computed using
the factorization model which describe rather well the measured branching
ratios for the  Cabibbo-allowed,
color-favored two-body $B$ decays to  charmed mesons \cite{Luo}. 
Writing

 $ A(B^- \to D^{*-}_s  D^{*0}) = 
\langle D^{*-}_s  D^{*0}|{\mathcal H}_{\rm eff}|B^{-}\rangle $ 
and 
\bea
&& \langle D_s^{*-} D^{*0} | {\mathcal H}_{\rm eff} | B^- \rangle =
\displaystyle{G_F \over \sqrt{2}}V_{cb}V_{cs}^* a_1 \nonumber \\
&& \times \langle D^{*0} | (V-A)^\mu | B^- \rangle
\langle D_s^{*-}| (V-A)_\mu | 0 \rangle
\label{fact2}
\eea
with
\bea 
<0|\bar q_a \gamma^\mu \gamma_5 c|D_a(v)> &=& f_{D_a} m_{D_a} v^\mu 
\nonumber \\
<0|\bar q_a \gamma^\mu c|D^*_a(v,\epsilon)> &=&f_{D^*_a} m_{D^*_a}
\eea
and
\bea
<D^0(v^\prime)|V^\mu|B^-(v)>&=&\sqrt{m_B m_D} \; \xi(v \cdot v^\prime)
(v+v^\prime)^\mu\nonumber \\
<D^{*0}(v^\prime,\epsilon)|V^\mu|B^-(v)>&=& - i
\sqrt{m_B m_{D^*}} \; \xi(v \cdot v^\prime)\nonumber \\
 &&\times  \epsilon^*_{\beta} \;
\varepsilon^{\alpha \beta \gamma \mu} v_\alpha v^\prime_\gamma \kern -0.7cm 
\label{BD} \\
<D^{*0}(v^\prime,\epsilon)|A^\mu|B^-(v)>&=&
\sqrt{m_B m_{D^*}} \; \xi(v \cdot v^\prime) \epsilon^*_{\beta}\nonumber \\
&&\times  
[(1+v \cdot v^\prime) g^{\beta \mu}- v^\beta v^{\prime\mu}] \nonumber
\eea
We next evaluate the scattering amplitude 
 $ A(D^{*-}_s  D^{*0} \to \chi_{c0} K^{-}) $ using the
$t-$channel $D^{0}, D^{*0}$ exchange Born term shown in diagrams of Fig.3.

The couplings $D D^{*} K$ and $D^{*} D^{*} K$ for the strong vertex
in the scattering amplitude can be
expressed in terms of a parameter $g$ in HQET\cite{Wise} and can be 
extracted from the $D D^{*} \pi$ coupling
using $SU(3)$ and extrapolated from the soft pion limit. 
\bea
<D^0(p) K^-(q)|D_s^{*-} (p+q,\epsilon))>&=&
g_{D_s^{*-} D^0 K^-} \, \, (\epsilon \cdot q) \nonumber\\
<D^{*0}(p,\eta) K^-(q)|D_s^{*-} (p+q,\epsilon))>\kern -0.2cm 
&=&\kern -0.2cm i 
\epsilon^{\alpha \beta \mu \gamma} \, p_\alpha \epsilon_\beta
\, q_\mu \eta^*_\gamma \nonumber\\
&& \times g_{D_s^{*-} D^{*0} K^-}  \label{gddk}
\eea
with 
\be
g_{D_s^{*-} D^{0} K^-}= -2\frac{\sqrt{m_D m_{D_s^*}}\,g }{f_K} ,
g_{D_s^{*-} D^{^*0} K^-}= 2\frac{\sqrt{m_{D_s^*} m_{D^*}}\,g }{f_K}
\label{gddk2}
\ee
 As $g$ is defined as the coupling in the decay
$D^{*}\to D\pi$, its soft pion value is known from experiment which gives
$g=0.59 \pm 0.01\pm 0.07$. This value supports  a previous prediction
using soft pion theorem which gives $g=0.75$ \cite{Pham}.  
We include off-shell effects for each 
strong vertex with coupling $g_{i}$ by a
form factor taken as \cite{Gortchakov,Navarra}  
$F_{i}(t)=( \Lambda^{2}_{i} - m^{2}_{D*})/(\Lambda^{2}_{i} - t)$. 
These form factors also act as suppression factor for the charmed meson
rescattering ammplitudes which, being exclusive process at high energy
(in the $B$ mass region), should be suppressed. 
For the   $\chi_{c0} D D$ and $\chi_{c0} D^{*} D^{*} $ vertex, as with
the $D^{*} D K$ vertex, we extrapolate the  on-shell couplings  
to off-shell $t$ region with a form factor similar to $F_{i}(t)$. 
The on-shell couplings  are obtained assuming the dominance 
by the nearest scalar meson state for the 
scalar $\bar{c}\,c$ current. The couplings are then
\be
g_{\chi_{c0} DD}\kern -0.1cm =\ \kern -0.1cm - 2  { m_D m_{\chi_{c0}}
  \over f_{\chi_{c0}}},\quad  
g_{\chi_{c0} D^* D^*}\kern -0.1cm =\ \kern -0.1cm  2  { m_{D^*} m_{\chi_{c0}}
\over f_{\chi_{c0}}} 
\label{gchi}
\ee
with 
$f_{\chi_{c0}}= 519\pm 40\rm\,MeV $ 
from QCD sum rules \cite{Colangelo} . Similarly, the       
$J/\psi D D$ and $J/\psi D^{*} D^{*} $ couplings are 
obtained with $J/\psi$ dominance for the
charm quark contribution to the electromagnetic form factor of the $D$
meson at zero momentum transfer~. We find 
\be
g^{2}_{J/\psi D D}/ 4\,\pi=5. 
\label{DDpsi}
\ee
though large. but not as large as the value 
$g^{2}_{\psi(3770) D D}/ 4\,\pi= 17.5$ obtained from the 
width of the $\psi(3770)$. For comparison,  
$g^{2}_{\phi K^{+} K^{-}}/ 4\,\pi= 1.66$.
Thus there is evidence that  charmonium $1^{--}$ states  couple strongly
to charmed mesons \cite{Achasov}. One consequence of the large coupling  
$g_{\psi(3770) D D}$ is that the rescattering effects would be more
important in $B^{-}\to \psi(3770) K^{-}$. The small
leptonic decay constant of the $\psi(3770) $ furthermore
makes the factorizable term less significant so that most of 
the contribution wouild come from the non factorizable terms.
Thus one should see the decay 
$B^{-}\to \psi(3770) K^{-}$ with a branching ratio comparable to 
 $B^{-}\to \chi_{c0} K^{-}$ branching ratio.

\section{Results}
In  Table 1, we give  the absorptive $\rm Im A$ and dispersive 
part $\rm Re A$ for the rescattering contributions.
\begin{table}[ht]
\caption{Numerical results for the rescattering contribution in
$10^{-7}\rm\,GeV$ ($A$) and in $10^{-7}$ ($\tilde A$)}
\label{table}
\begin{center}
\begin{tabular}{|| l || c c  || c |} \hline \hline
\kern -0.2cm $B^-\kern -0.1cm \to  \kern -0.1cm K^-\chi_{c0}$\kern -0.2cm &${\rm Re}\, {A}$ \kern -0.3cm&${\rm Im}\, {A}$ 
& $\Lambda_i$ (GeV)\kern -0.2cm \\\hline
\kern -0.2cm&$-(0.9-1.7)$\kern -0.3cm&$-(0.5-1.0)$&$2.5$\\
\kern -0.2cm&$-(1.4-2.7)$\kern -0.3cm&$-(0.6-1.2)$& $2.8$\\
\hline \hline
\kern -0.2cm $B^-\kern -0.1cm \to \kern -0.1cm K^-J/\psi$\kern -0.2cm &${\rm Re} \,\tilde {A}$\kern -0.3cm&${\rm Im} \,\tilde {A}$&$\Lambda_i$ (GeV)\kern -0.2cm\\ \hline
&$(0.1-0.2)$\kern -0.3cm&$-(0.5-0.9)$& $2.5$\\
&$(0.2-0.3)$\kern -0.3cm&$-(0.9-1.7)$&$2.8$\\
\hline \hline
\end{tabular}
\end{center}
\end{table}

The experimental  amplitudes are 
\bea
|A_{\rm exp}| &=& (3.39\pm 0.68)\times 10^{-7}\rm\,GeV , \quad
(\rm Belle) \nonumber\\
& =& (2.1\pm 0.3)\times 10^{-7}\rm\,GeV , \qquad  (\rm Babar)
\nonumber
\label{Aexp}
\eea
for  $B^{-} \to \chi_{c0} K^{-}$. For $B^{-} \to J/\psi K^{-}$, we have:
\be
|\,\tilde A_{\rm exp}| = (1.41\pm 0.68)\times 10^{-7} , \quad (\rm PDG)
\label{A1exp}
\ee
where $\,\tilde A$ defined as
\be
A(B^{-} \to J/\psi K^{-}) = \tilde A\,\epsilon^{*}\cdot q
\label{A1}
\ee
We see from Table 1 that both the real and imaginary parts of the $B^{-} \to
\chi_{c0} K^{-}$ and $B^{-}\to J/\psi K^{-} $ decay amplitudes are 
in the range of the measured values.

\section{Conclusion}
Charmed meson rescattering seems capable of explaining the large
branching ratio for the decay $B^{-} \to \chi_{c0} K^{-}$. It also 
makes an important contribution to the $B^{-}\to J/\psi K^{-} $ 
decay rate which would be too small in QCD factorization. The recent
observation of the decay $B^{+}\to \psi(3770) K^{+} $ at Belle \cite{Abe}
with a branching ratio of $(0.48\pm 0.11\pm 0.12)\times 10^{-3}$
could be another strong evidence of non factorizable terms in 
non leptonic $B$ decay with charmonium in the final state. One could
look for more evidence \cite{Pham1} in other $B$ decays to $P-$wave
 charmonium states such as the $\chi_{c2}$ and the $h_{c}$ meson 
which have no visible factorizable contributions.

\section*{Acknowledgments}
I would like to thank G. Nardulli, P. Colangelo and the Organisers 
of the QCD@Work 2003  Workshop for the warm hospitality extended to me
at Conversano.


\begin{thebibliography}{9}

\bibitem{Colangelo} P. Colangelo, F. de Fazio  and T. N. Pham,
Phys. Lett. {\bf B} 542 (2002) 71.

\bibitem{Belle1} K. Abe {\em et al.}, Phys. Rev. Lett. {\bf  88} 
(2002) 031802.

\bibitem{Babar} B. Aubert et al., hep-ex/0207066.

\bibitem{PDG} Particle Data Group, K. Hagiwara {\em et al.},
Phys. Rev. D {\bf 66}  (2002) 010001-1.

\bibitem{Suzuki} M. Suzuki,
Phys. Rev. D {\bf 66}  (2002) 037503.

\bibitem{Isola} C. Isola, M. Ladisa, G. Nardulli, T. N. Pham, and 
P. Santorelli, Phys. Rev. D {\bf 64}  (2001) 014029,  
ibid.,  Phys. Rev. D {\bf 65} (2001) 014005 .

\bibitem{Kamal} A. N. Kamal, Int. J. Mod. Phys. A  {\bf 7} (1992) 3515 .

\bibitem{Achasov} N. N. Achasov and A. A. Kozhevnikov, 
Phys. Rev. D {\bf 49} (1994) 275.

\bibitem{Choudhury} D. Choudhury and J. Ellis, Phys. Lett. {\bf B} 433 
(1998) 102.

\bibitem{Colangelo1} P. Colangelo, F. de Fazio P. Santorelli and
E. Scrimieri, Phys. Rev. D {\bf 53} (1996) 3672~

\bibitem{Deshpande} N. G. Deshpande and X. G. He, Phys. Lett. 
B {\bf 336} (1994), 471; N. G. Deshpande, X. G. He, W. S. Hou, and S. Pakvasa,
 Phys. Rev. Let. {\bf 82} (1999), 2240.

\bibitem{Isola1} C. Isola and T. N. Pham, Phys. Rev. D {\bf 62}  
(2000), 094002.  

\bibitem{Cheng} H. Y. Cheng and K. C. Yang,  Phys. Rev. D {\bf 59} 
(1999) 092004~) 

\bibitem{Chay} J. Chay and C. Kim, hep-ph/0009244.

\bibitem{Cheng1} H. Y. Cheng and K. C. Yang, 
Phys. Rev. D {\bf 63} (2000) 074011~.  

\bibitem{Beneke} M. Beneke, G. Buchalla, M. Neubert, and
C. T. Sachrajda, Nucl. Phys. {\bf 606} (2001) 245~

\bibitem{Song} Z. Song and K. T. Chao,
Phys. Lett. {\bf B} 568 (2003) 127.

\bibitem{Luo} Z. Luo and J. L. Rosner, 
Phys. Rev. D {\bf 64} (2001) 094001~) 

\bibitem{Wise} M. B. Wise . Phys. Rev. D {\bf 45} (1992) 2188;
G.~Burdman and J.~F.~Donoghue, Phys. Lett. B
{\bf 280} (1992) 287; T.~M.~Yan {\em et al.}, Phys. Rev. D {\bf 46}
(1992) 1148).

\bibitem{Pham} T. N. Pham, Phys. Rev. D {\bf 25} (1982) 2955.

\bibitem{Gortchakov} O. Gortchakov, M. P. Locher, V. E. Markushin, and
S. von Rotz, Z. Phys. {\bf A 353} (1996) 447.

\bibitem{Navarra} F. S. Navarra, M. Nielsen, and M. E. Bracco,
Phys. Rev. D {\bf 65} (2002) 037502~). 

\bibitem{Abe} K. Abe {\em et al}, Belle Collaboration,
contributed paper at the 2003 International Europhysics Conference 
on High Energy Physics and the XXI International Symposium on Lepton 
and Photon Interactions at \break High Energies, hep-ex/0307061.

\bibitem{Pham1} P. Colangelo, F. de Fazio  and T. N. Pham,
hep-ph/0310084.

\end{thebibliography}
\end{document}